# Probing the membrane potential of living cells by dielectric spectroscopy

Corina Bot and Camelia Prodan

Dept. of Physics, New Jersey Inst. of Techn, Newark, NJ

**Abstract.** In this paper we demonstrate a quantitative way to measure the membrane potential of live cells by dielectric spectroscopy. We also show that the values of the membrane potential obtained using our technique are in good agreement with those obtained using traditional methods-voltage sensitive dyes. The membrane potential is determined by fitting the experimental dielectric dispersion curves with the dispersion curves obtain from a theoretical model. Variations in the membrane potential were induced by modifying the concentration of potassium chloride in the solution of the cell suspension in the presence of valinomycin. For exemplification of the method, E. coli were chosen for our experiments.

**Introduction**

The resting membrane potential is one of the most important living cell parameters. By monitoring a cell's membrane potential, one can get important information about its state. Changes in the membrane potential have been linked to different diseases, including Parkinson's, epilepsy and Bartter's syndrome [1]. A decrease in the plasma membrane potential has been seen in tumor cells with increased expression of multi-drug resistance (MDR) protein, also known as P-glycoprotein [2]. It was also shown that the membrane potential plays a critical role in the antibiotic uptake by cells and in bactericidal action [3]. Thus measuring or monitoring the membrane potential is of great importance. The existing "gold" or standard methods to measure or monitor the membrane potential are patch clamping and voltage sensitive dyes. In patch clamping, one measures the absolute value of the membrane potential while with voltage sensitive dyes one can only monitor changes in the membrane potential. Although both methods are very effective, they can be invasive or laborious.
Dielectric spectroscopy (DS) studies the complex dielectric function of the biological samples, which can be cells suspended in solution or tissues. Because it is noninvasive, this method is widely used in many areas including biophysics and pharmacology. For example DS has been used to monitor the living cell and plasma membrane properties or protein activity [4-7]. Impedance spectroscopy (IS) was also used to characterize the wound healing process of monolayer cell tissues [8]. The single cell DNA content, for example, was reported to be determined using impedance monitoring [9]. In pharmacology, cellular dielectric spectroscopy has been used as a label free technology for drug discovery [5]. Low frequency IS and DS are well-established methods to study live cells such as bacteria or mammalian cells. Pathogens frequently have complex biological matrices but they lack the chemical functional groups on their cell walls, thus leading to difficulties in designing biosensors for their detection. IS has proved to be a very useful technique in early detection of bacterial colonies, allowing precise monitoring of the cellular behavior in a noninvasive manner [10-16]. Other researchers have quantified a detection time for Salmonella typhimurium and E. coli O157:H7 cultures from the impedance growth curve (impedance against bacterial growth time) at 10 Hz [17]. Also, rod-shaped cells (a Streptomyces strain with a filamentous cellular shape) presented in a study [18] were measured in the 10Hz -13MHz frequency range. Studies involving E. coli cells showed mainly the beta dispersion effects, in the frequency range of: 20kHz to 20MHz [19], 75kHz to 20MHz [20] and 1kHz to 110 MHz [21]. The dielectric measurements of an Escherichia coli cell suspension and liposome suspensions were carried out in the frequency range between 0.1 and 100 MHz to detect the heat stress-mediated interaction between proteins and cell membranes [22]. Recently, dielectric measurements on mammalian cells present dielectric permittivity screening results for cytoplasm and membrane specific markers in the frequency range 1 kHz to 100 MHz, using different intracellular proteins [23].

It was previously shown that the presence of the membrane potential has a specific effect on the dielectric behavior of cell suspensions, namely the appearance of a frequency dependence called the α-dispersion, which is present in the low frequency part of the dielectric dispersion curves [24, 25]. Thus, low frequency dielectric spectroscopy could be developed into a label-free, fast and noninvasive technique to measure or monitor the membrane potential and its changes. This represents the main purpose of this article.

In this paper we demonstrate a quantitative way to measure the membrane potential of living cells by dielectric spectroscopy which is applicable to any type of live cells. We also show that the values of the membrane potential obtained using our technique are in good agreement with those obtained using traditional methods-voltage sensitive dyes. For the exemplification of the method, E. coli was chosen for our experiments.

The membrane potential of E. coli is determined by fitting the experimental dielectric dispersion curves with a theoretical model [24, 25]. Variations in the membrane potential were induced by modifying the concentration of potassium chloride in the solution in the presence of valinomycin. The variations in the membrane potential were quantified using DS and the traditional method of voltage sensitive dyes. The cells were prepared in the same way for both types of experiments. For DS measurements, we applied a recently introduced method to remove the electrode polarization error from the live cell suspension dispersion curves [26].

**Material and methods**

E. coli K12 wild type (www.atcc.com) were incubated in 5 ml of Tryptone Soy broth (growth media) at 175 rpm and 37°C to the saturation phase for 16 hours. For cells viability in water, the culture was then resuspended in 60 ml of media and incubated in the same conditions to a mid logarithmic phase, for 4 hours. For dielectric spectroscopy measurements, cells were incubated overnight in 3 test tubes, each with 5 ml of Tryptone Soy broth then mixed together. After reaching saturation phase, 2 ml of cells were resuspended in 23 ml of media and incubated in the same conditions to a mid logarithmic phase, for 2.5 hours to an OD of 0.615.

For cell viability in water, two test tubes with 10 ml of cells in media and cells in autoclaved DI water were tested for 7 days. Each day, cells were cultivated on agar plates by serial dilutions and incubated for 18 hours at 37°C and the number of colonies counted from the $10^{-7}$ area. In parallel, a test tube with cells in media was kept in the same conditions. Every day a serial dilution was performed and plates were inoculated for each cell concentration. Cells were imaged at 100x magnification and pictures were taken with a CCD camera (Sensicam qe The Cooke Corporation) attached to a Zeiss Axiovert 200 inverted microscope.

For DS, a fresh suspension of cells was centrifuged at 2522xg for 6 min before each measurement and the pellet resuspended in ultrapure water (from a Milli-Q Direct Q water system) with 5mM glucose to an $OD_{600}$ of 0.317. Glucose was used to balance the osmotic pressure.

Fluorescence measurements were performed on cells harvested and resuspended under the same conditions as for dielectric spectroscopy with a Varian–Cary Eclipse

Spectrophotometer. The voltage sensitive dye used for staining the cell membrane was Di-SC$_3$-5 Carbocyanine iodide (www.anaspec.com) with an excitation wavelength of 642 nm and emission wavelength 670 nm. Dye was added in a concentration of 0.4 μM [27] both to cells in media and cells in water. All fluorescence measurements were performed after a cell incubation of 15 min, to allow the dye to settle in the membrane.

Dielectric spectroscopy measurements were performed with a Two-Channel Dynamic Signal Analyzer SR785 over the frequency range 10 Hz to 100kHz and with a Solartron 1260 Impedance Analyzer over the frequency range 10 Hz to 1MHz. We use the same experimental setup as previously described in Ref. [28]. The cell suspension to be measured is placed between two parallel gold-plated electrodes that are enclosed in a cylindrical glass tube. The distance between the capacitor plates is controlled with a micrometer. The results reported here were obtained with electrodes of 28.1 mm in radius and an applied electric field of 0.3V. The distance between the electrode plates was 3mm. The setup provides the complex impedance Z of the measuring cell over the frequency ranges mentioned above.

A major source of error when measuring the impedance of liquid samples is the polarization that appears at the contacts between the sample and electrodes [28]. The superficial layer of charges near the electrodes polarizes when electric fields are applied, leading to the so-called electrode polarization effect, which is predominant at low frequencies. An efficient method for polarization effect removal was developed in Ref. [26], which we apply in the present work. The measured impedance Z is actually a sum of the intrinsic impedance of the sample and the impedance of the polarization layers: $Z=Z_s+Z_p$. The polarization impedance is caused by the medium in which the cells are suspended rather than the cells themselves. Thus, the polarization impedance can be obtained from measurements on the medium alone. Hence, after each measurement, the sample was centrifuged and the supernatant was collected from each sample. Extreme care was paid to minimize possible breaking of the cells. For the first set of measurements, cells were centrifuged at 2522xg and then supernatant was collected. The solution with cells at OD 0.206 was centrifuged for 10 min and then the supernatant was collected. For OD 0.434 and OD 0.638 cells were centrifuged 2 min, the supernatant was collected and then centrifuged again for 2 min, and the same procedure repeated for another 4 min. The DS measurements on the resulting supernatant (centrifuged 10 min total) were done in the same conditions as the cells. For the higher concentration of cells (i.e., OD 0.4, 0.6), a much gentle centrifugation is required to avoid membrane breakage. The membrane breakage releases ions, hence increasing the conductivity of the supernatant. To be absolutely sure that the supernatant had same ionic properties as the medium in which the cells were suspended during the DS measurements, high frequency conductivity measurements were taken from the cell suspension and from the supernatant, making sure that the two values coincided.

Once the medium was separated, we eliminated $Z_p$ by applying the following steps.
i) Record the impedance of the medium $Z_m$ for each frequency.
ii) In the high frequency range, where the polarization effect is absent, fit $Z_m$ with the ideal impedance given in equation (1), and use the dielectric permittivity ε and conductivity σ as fitting parameters:

$$Z_{ideal} = \frac{d/A}{j\omega\varepsilon + \sigma} \qquad (1)$$

In this equation, d represents the distance between the capacitor plates and A their surface area.
ii) Once the intrinsic dielectric constant ε and conductivity σ of the medium was obtained, use these values to extrapolate $Z_{ideal}$, which now represents the intrinsic impedance of the medium, all the way to zero frequency.
iii) Compute $Z_p$ as the difference between $Z_m$ and the extrapolated $Z_{ideal}$.
iv) Correct the impedance Z by subtracting $Z_p$ calculated above and obtain the intrinsic impedance of the sample $Z_s$.

For fluorescence and dielectric measurements, 1µM valinomycin was added [27] to make the membrane permeable to $K^+$. Valinomycin was bought from Molecular Probes and hydrated in DMSO spectrophotometric grade, 99.9% (ACROS Organics) to a 2.2mM stock concentration. Valinomycin is a natural electrically neutral ionophore, which is highly selective for potassium ions over sodium ions within the cell membrane [27, 29-33]. It functions as a potassium-specific transporter and facilitates the movement of potassium ions through lipid membranes by following an electrochemical potential gradient. Previous work has shown that valinomycin replaces the hydration component of complexed ions with a hydrophobic coat, and this results in a biphasic current-time course on its interaction with the lipid bilayer [32].

**Results**

In order to quantify the intrinsic response of the cells, it is desirable to have an electrode polarization effect as low as possible. For this reason, one would like to consider a suspension fluid with a low ionic strength and this motivated our choice for pure water with 5mM glucose as suspension media for the E. coli cells. Before the actual dielectric measurements, we tested the viability of the cells in sterile DI water over a period of seven days as described in "Materials and Methods". In Fig. 1 we present the results of this viability study. From each inoculated plate, the number of colonies from $10^{-7}$ dilution was counted and plotted vs. time. In the first 100 hours, we can identify a slight growth followed by a regression; also, from 150 hours on, the same pattern is followed. This slight growth is expected since some cells follow apoptosis, breaking down and releasing nutrients for the other living cells.

Next, we wanted to study how the cell membrane potential responds to the transition from growth media to the suspending media, water with glucose, by means of voltage sensitive dyes. We observed that, when the cells are harvested from the growth media and placed in water, their membrane potential changes. This is a direct result of their adaptability to the new conditions. Fluorescence intensity of cells in media and cells in water (average of 5 measurements) is plotted vs. wavelength in the inset in Fig.1. The cells in water exhibit a decrease in fluorescence intensity, which is directly proportional to a decrease in membrane potential (hyperpolarization). Hyperpolarization occurs first due to a ph change (from acidic to neutral) and second due to a hyperpolarization of the cell membrane when $K^+$ diffuses outside to balance Δψ to equilibrium. In glucose-containing media, E. coli ferments glucose to produce lactic, acetic and succinic acids. Not only do these products decrease the pH of the medium, they also increase admittance values of the cell suspensions [15].

After these tests, which showed that the cells are viable in water with 5mM glucose, we performed DS measurements on different cells concentrations, and we correlated them to optical density recordings (OD). The methodology was described in "Materials and Methods" section. A thorough analysis of the imaginary part of the impedance is required in order to get a good dielectric characterization for each measurement.

In Fig. 2 a, b and c we present the imaginary part of: the measured (un-corrected) impedance of cell suspension Im[Z] (full circles), the impedance of the medium or supernatant

Im[$Z_m$] (solid line), the polarization impedance Im[$Z_p$] (dashed line) and finally the corrected or the intrinsic impedance of the cell suspension for three concentrations (OD), obtained after polarization removal Im[$Z_s$] (open circles). As one can see, Im[$Z_p$] becomes negligible starting with 10 kHz for each cell suspension. In other experimental trials (not shown here), if the cells were not centrifuged gently, the conductivity of the supernatant became much larger than the conductivity of the cell suspension, due to cell lysis. This conductivity issue is directly seen in an increase in Im Z of the supernatant. In other words, the conductivity of the cells and collected supernatant must be very carefully monitored and their values should be similar (i.e. from 0.0030 to 0.0035 S/m).

The polarization effect increased the relative dielectric permittivity of the suspension at 10 Hz on average by approximately half of an order of magnitude. The inset in Fig. 2b shows a magnification of the plots, which reveals that that Im[Z] cells is significantly different from Im[$Z_m$] of the supernatant. Conductivity of the supernatant was also measured at 25°C with a conductivity cell (Denver Instrument 220 Ph Conductivity Meter). The measured and the fitted conductivities are in agreement to the first significant digit. The fitting parameters ε and σ for the medium are presented in Table 1, along with the measured conductivity of the supernatant.

Table 1:

| O.D. | ε fitted | ε fitted % error | σ fitted (S/m) | σ fitted % error | σ measured (S/m) @ 25.8°C | R |
|---|---|---|---|---|---|---|
| 0.206 | 72.565 | 0.26791 | 0.0030821 | $6.8273*10^{-6}$ | 0.00309 | 0.99871 |
| 0.434 | 70.713 | 0.1624 | 0.0057454 | $2.7707*10^{-6}$ | 0.00541 | 0.99921 |
| 0.638 | 67.129 | 0.23662 | 0.0097584 | $3.1789*10^{-6}$ | 0.00929 | 0.99817 |

In Fig. 2 a, b, and c, in the imaginary part of each of the corrected curves for various cells concentrations, we observe the appearance of a small plateau in the frequency region of 80 to 1000 Hz. In Fig. 2 a and b, the plateau is more like a prominence because the cell concentration is small, but in Fig. 2 c the plateau is clearly visible. This directly translates into small plateaus in the intrinsic dielectric curves of the cell suspensions; the higher the cell concentration, the more predominant the effect. This could be evidence of the beta plateau for E. coli cells.

The intrinsic relative dielectric permittivity for the three suspensions with different E. coli concentrations is plotted in Fig. 2 d. At low frequencies, the relative dielectric permittivity is increasing as the concentration of cells is increased (correlated with the optical density). The intrinsic relative dielectric permittivity for OD 0.434 has values of the order of $6.5 \times 10^5$ at low frequency and the values are decreasing towards 70 at high frequencies. The intrinsic relative dielectric permittivity for OD 0.638 has the highest value equal to $7.9 \times 10^5$, which occurs at low frequencies, and the values are decreasing towards 67 at high frequencies. As the optical density is increased, the conductivity of the cell suspension (and collected supernatant) also increases due to a larger depletion of salts as metabolites. Thus, the electrode polarization error increases with the increase in the cell concentration.

We now move on to the experiments in which we induced variations in the membrane potential. These variations were induced using KCl and in the presence of valinomycin. During the early logarithmic phase of growth, the intracellular K concentration in E. coli exceeds that of the external medium [34]. When valinomycin was added to the suspension, the cells' membranes were made permeable to the $K^+$ ions inside the cells and a flow started due to unhindered diffusion. Since $K^+$ is more concentrated inside the cells, the net movement of the charges is from inside out, down the concentration gradient, until an equilibrium membrane potential is reached. This is a common mechanism by which many cells re-establish a membrane potential.

In Fig. 3, the effect of valinomycin on the cellular membrane potential was probed using voltage sensitive dyes. Again, we see a decrease in the fluorescence intensity after 20 min of adding valinomycin and it proves the effort of the cells to balance $\Delta\psi$. At some point, an equilibrium resting membrane potential is reached. Further variations in the resting membrane potential were induced by changing the potassium concentration outside the cells. This will change the potassium concentration gradient and a new resting potential will be established. This can be directly correlated to the fluorescence quenching of Di-SC3-5, which is classified as a slow-response probe with changes in intensity of 0.2 - 0.5% per mV change in membrane potential [35-37]. In Fig. 4 we present the fluorescence intensity after the addition of KCl to the suspension of cells. The intensity plotted at 664 nm has an increasing trend when [μM] KCl is added outside the cells. These changes in intensity are proportional to a membrane depolarization and an increase in membrane potential. By calculating the percentage change in fluorescence as $\Delta F=100(F_f-F_i)/F_i$ [38] and using 0.5% per mV change [36] we obtained an average of 10mV membrane potential increase correlated to $\Delta F$. In the inset in Fig. 4, the raw data for the fluorescence quenching is presented as a function of wavelength.

In order to probe these changes in membrane potential using DS, we used suspensions of cells in ultrapure water with 5mM glucose, under the same conditions as for fluorescence measurements. Each suspension was accommodated with 1 μM valinomycin and after approximately 20 min, [μM] KCl was added externally. In Fig. 5 a, the relative dielectric permittivity computed from the raw data is plotted for two KCl concentrations compared with cells only; these were chosen because they exhibited a more significant change in fluorescence (as seen in Fig. 4). After the polarization errors are removed as described before, we can easily distinguish between the curves corresponding to various [μM] KCl added (see Fig. 5 b). In this latter figure, all the dispersion curves of the intrinsic relative dielectric permittivity are decreasing with increasing [μM] KCl, in accordance with the theoretical model described in Refs. [24, 25]. Indeed, with the addition of KCl, the cell membrane depolarizes and, according to the theoretical models, one should observe a decrease in the low frequency dielectric function.

The fitted parameters for the supernatant are presented in Table 2:

Table 2:

| [KCl] | ε fitted | ε fitted % error | σ fitted | σ fitted % error | R value |
|---|---|---|---|---|---|
| 0μM KCl | 69.152 | 0.22986 | 0.0036964 | $8.7789*10^{-8}$ | 0.99458 |
| 3μM KCl | 66.741 | 0.32505 | 0.0040152 | $1.1231*10^{-7}$ | 0.98882 |
| 5μM KCl | 68.012 | 0.26059 | 0.0036185 | $9.9382*10^{-8}$ | 0.99273 |

The error bars represent the standard deviation over 4 to 5 experiments for each dispersion curve.

**Discussion**

Currently used techniques for sterilizing drinking water from coliforms or other bacteria starvation assays [39] start with resuspending the cells in Milli-Q water without any source of energy. E. coli cells are usually left to settle with or without shaking overnight before any sterilization tests [40]. This motivated our interest in having fully viable cells in Milli-Q water with 5mM glucose and using them for measurements. Each measurement lasts less than 25 min, including the incubation time required by valinomycin.

The first set of measurements indicates that DS can provide an effective way to distinguish between cell suspensions with different volume fractions. In Fig. 2 d, we indicate the changes in the relative dielectric permittivity related to cell concentration. The lowest concentration had the lowest values for ε, and the increase in the cell concentration is directly proportional to the increase in ε. The parameters given by the fitting when removing the polarization effects and listed in Table 1 and Table 2 have a low error for ε (0.2%) and even lower for σ (~$10^{-7}$%). The experimental outcomes for the relative dielectric permittivity at high frequencies are in good agreement with the results presented by other authors, for example in Ref. [41], who found values ranging from 70 to 74 for suspensions of ellipsoidal cells. For E. coli in culture medium, Ref. [42] found ε values ranging from 92 to 95 at 5MHz and Ref. [43] had roughly 100.

The second set of experiments indicates that DS can distinguish between suspensions of cells with different membrane potentials. In these experiments, the concentration of cells (OD) was kept the same; just the membrane potential was varied from one experiment to another. The membrane potential was changed in steps of 10mV using KCl. Using images taken with the microscope, we concluded that the addition of the KCl did not change the shape or volume of the cells, due to osmotic pressure.

Next, our experimental results for the dielectric permittivity are combined with the theoretical model from [24] in order to obtain the value of membrane potential from the DS measurements. It was recently shown [25] that this theoretical model has an analytic solution if the cells are considered spherical. For simplicity, we will work with this analytic model and approximate E. coli cells with spherical cells as best we can. The independent input parameters for the model are the following: The volume fraction of the suspension; The size (radius) of cells and the thickness of the membrane, which we fixed at 1.5 μm and 2 nm, respectively; The dielectric constant and conductivity of the medium, which we already listed in Tabels 1 and 2; The dielectric constant of the membrane, which we fixed at a value of 34 (from Ref. [43]); The dielectric constant and conductivity of the inner cell region, which we fixed at 78 and 0.2 S/m (from Ref. [45]); The diffusion constants of surface charges accumulated at the outer and inner surfaces of the membrane, which were adjusted for the best fit of the data. The last input parameter for the model is the membrane potential. Using the analytic model, we have simulated in Fig. 6 a and b the experimental dielectric curves reported in Fig. 2 d, as well as those reported in Fig. 5 a and b.

For the first set of data, the volume fraction considered was 0.015 for OD 0.206, 0.06 for OD 0.434 and 0.09 for OD 0.638; these values were obtained based on images of cells taken using the Zeiss Axiovert microscope and the camera described before. A resting membrane potential of -180 mV was kept the same for all three concentrations. We would like to comment on this value. Cells in media undergoing pH changes from acidic to alkaline normally have a resting membrane potential of -150 mV [44]. When cells are harvested from the growth media and placed in water with 5mM glucose, they hyperpolarize as shown in Fig. 3. Moreover, the addition of valinomycin hyperpolarizes the membrane by another 10 mV, which adds over the hyperpolarization that cells undergo when they are resuspended in water and glucose. Thus, using the fluorescence measurements, we estimated that the membrane potential of cells in water with 5mM glucose would be around -180mV. The results of this model for cell concentrations are presented in Fig.6 a., where the solid lines represent the theoretical curves and the dotted lines are the experimental results.

For the second set of data presented in Fig. 5 a, where we studied the effect of membrane potential variations, the simulation is shown Fig. 6 b. The volume fraction for these suspensions was 0.02 (as obtained by direct imaging). The cytoplasm's conductivity was varied from 0.2 to 0.25 S/m because the conductivity inside the cells should increase with the addition of KCl. The membrane potential was varied by depolarization in 10mV steps from -180 mV to -160 mV. In Fig. 6 b. the solid lines represent the theoretical curves and the dotted lines are the experimental results.

The experimental data shown in Figs. 5 and 6 show only one dispersion. The question we will address next is which plateau, alpha or beta? From previous experiments [43] we know that E. coli suspensions can exhibit a beta dispersion in the kHz-MHz region. A significant difference between our measurements and theirs is that in the latest the cells were measured in the growth media, which has a conductivity of about 440 times higher than the water, which was the medium for our experiments. To see what is the effect of such difference, we increased the conductivity of the medium from 0.006 to 0.03 S/m in our simulations and we plotted the results in Fig. 7. In these theoretical simulations we used a volume fraction of 0.09 and a resting membrane potential of -180mV; all other parameters were kept as before. For the curves shown in Fig. 7, the only parameter that was varied was the conductivity of the medium. One can observe, as the conductivity of the medium is increased, the beta plateau becomes more prominent and it extends in the region 10kHz to 1MHz. With a conductivity of 0.03 S/m, the theoretical results are in very good agreement with the experiments of Ref. [43] and now we can be sure that the beta plateau is absent in our experimental data because the conductivity of the medium was very low. Moreover, our theoretical simulations further indicate that the plateau seen in our data represents the alpha plateau since there is a strong variation with the membrane potential. The tails of the dielectric curves come together in high frequency (starting 3kHz) and have values close to the dielectric permittivity of water.

In this paragraph we roughly estimate the necessary concentration of KCl for a 10mV change in the membrane potential. The capacitive current $I_c$ that flows in or out of a capacitor at a single point to produce a change in membrane potential of amplitude dV, is given by the equation: $C_m \frac{dV}{dt} = I_c$ [45-47]. For the overall effect of the $I_c$, we need to consider the complete surface area A of an E. coli cell: $C_m \frac{dV}{dt} A = -I_c$. The cell membrane surface area of 1.767 x $10^{-11}$ $m^2$ and volume of 4.4167 x $10^{-18}$ $m^3$ were calculated as for a cylinder topped at both ends by two hemispheres, with radius of 0.75μm and overall length 3μm. These sizes are determined from images of cells suspended under the same conditions as for measurements and they are slightly higher than average values reported previously [48]. The cell membrane capacitance $C_m$ is 0.043F/$m^2$ [43]. A very rough calculation of an action potential upstroke of 10mV in 2 ms gives an $I_c$ of 0.379 x $10^{-11}$ C/s. By using this result and the elementary electron charge in Coulombs, we can estimate that 47412 charges flow across the membrane in 2ms to change Δψ by 10 mV. The potassium concentration that corresponds to this flow is calculated by dividing the number of charges to Avogadro's number x volume of one cell, and is approximately 17μM. This is just a rough approximation, but by more elaborate calculations at a molecular level we can find this concentration more accurately. Based on the same approximation, and on the fact that valinomycin creates pores in the membrane that allow a free flow of the $K^+$ ions from the higher concentration towards the lowest, if we only change the time interval in the calculations to 0.02 s or even to 0.2 s there is no significant difference in the calculated [μM] $K^+$ concentration.

**Conclusions**

In this paper we demonstrate a quantitative way to measure the membrane potential of living cells by dielectric spectroscopy. This method could be applied to any type of live cells in suspensions. To obtain the membrane potential, we fitted out experimental dispersion curves with the ones obtained from a theoretical model, with the membrane potential as the fitting parameter. We also show that the values of the membrane potential obtained using our technique are in good agreement with those obtained using traditional methods-voltage sensitive dyes.

To obtain the intrinsic dispersion curves for cell suspensions, the polarization error was removed by using the method from Ref. [26].

Electrical impedance measurements were made on a culture of E. coli cells in the frequency range of 10 Hz-1MHz using a two-electrode system. First, the measurements made on three different concentrations of cells in suspensions using this technique show that the increase in dielectric permittivity after polarization removal is proportional to changes in optical density, thus to the concentration of cells. In the alpha frequency domain, the lowest concentration of cells in the suspension had the lowest values for ε, and an increase in cell concentration is directly proportional to an increase in ε. In the high frequency domain (10 kHz and up) the relative dielectric permittivity is close to that of the suspending medium and the values are in the range of 67 to 72 for various cells concentrations. Second, we presented the relationship between membrane potential variations and dielectric permittivity in the low frequency range. Experimentally, the membrane potential was changed using different concentrations of KCl in the presence of valinomycin. KCl depolarizes the cell membrane, and this was observed as a direct decrease in the low frequency dielectric constant (alpha relaxation) for each KCl concentration considered. By using the theoretical model presented in Ref. [24, 25] we extracted the membrane potential for each KCl concentration. Conductivity of the supernatant needs to be closely monitored, as it plays a major role in removing the polarization error (i.e. if the cells are centrifuged gently, the conductivity of the supernatant is not unnecessarily increased due to cell lysis).

In conclusion, we presented here a method to obtain and monitor the membrane potential by dielectric spectroscopy, which is noninvasive and label free. This method could be used for fundamental research in life sciences as well as developed into a high throughput screening for pharmaceutical agents.

**Acknowledgment**:

This work was supported, in part, by the grant NJIT-ADVANCE awarded by the National Science Foundation (grant nr.0547427).

**Figure captions:**

Fig. 1 Number of E. coli colonies in water and media counted from $10^{-7}$ area of serial dilution vs. time. Solid columns represent cells in media and dashed columns cells in water. Inset represents fluorescence intensity changes vs. wavelength for cells in media (solid circles) and cells in water (line)

Fig. 2 a., b., c., Imaginary parts of impedances: cell suspension (Im $Z_s$ - circles), supernatant (solid line), polarization error (Im $Z_p$ - dashed line) and corrected cells after polarization removal (empty circles) plotted as a function of frequency. d. Corrected relative dielectric permittivity for the three suspensions with different E. coli concentrations OD 0.206 (dotted line), 0.434 (empty circles), 0.638 (solid line)

Fig. 3 Fluorescence intensity vs. wavelength for cells in water: cells only (circles), cells after 7 min of valinomycin addition (solid line), cells after 15 min of valinomycin addition (triangles)

Fig. 4 The fluorescence intensity plotted at 664 nm with the addition of KCl to the suspension of cells. Inset: raw data for the fluorescence quenching as a function of wavelength.

Fig. 5 a. Relative dielectric permittivity of E. coli and [µM] KCl computed from the raw data; b. Relative dielectric permittivity of E. coli and [µM] KCl after polarization removal.

Fig. 6 a. Relative dielectric permittivity of E. coli for OD 0.206 (triangles), 0.434 (squares), 0.638 (circles): theoretical simulations (empty symbols) and experimental results (full symbols); b. relative dielectric permittivity of E. coli for membrane potential: -180 mV (triangles), -170 mV (squares), -160 mV (circles): theoretical simulations (empty symbols) and experimental results (full symbols);

Fig. 7 Relative dielectric permittivity of E. coli vs. frequency: theoretical simulations for outside medium conductivity ($\sigma_o$) variations ($\sigma_o$ = 0.006 S/m dotted line, $\sigma_o$ = 0.01 S/m solid line, $\sigma_o$ = 0.02 S/m empty circles, $\sigma_o$ = 0.03 S/m empty triangles)

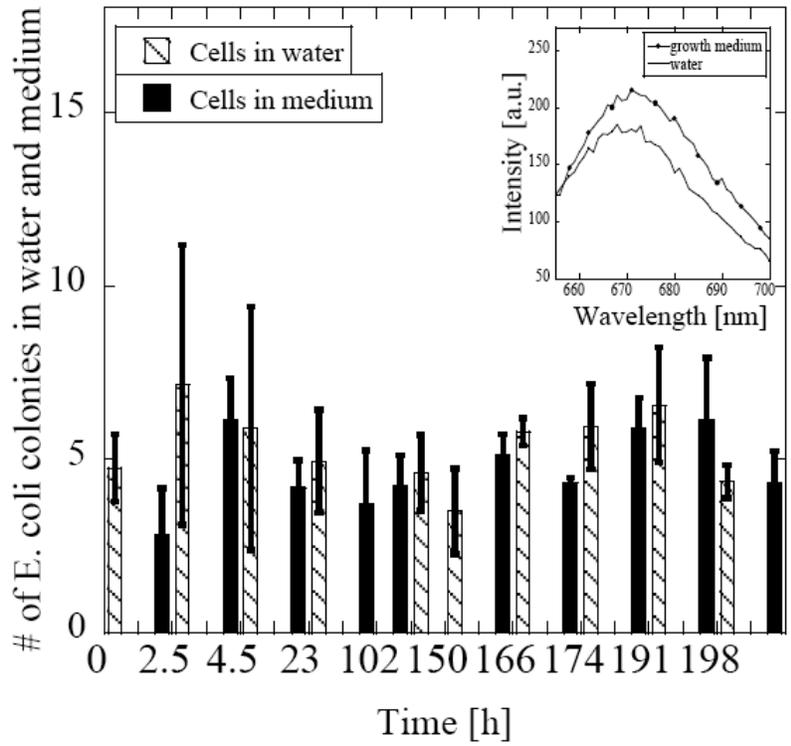

Fig. 1 Bot et al 2008

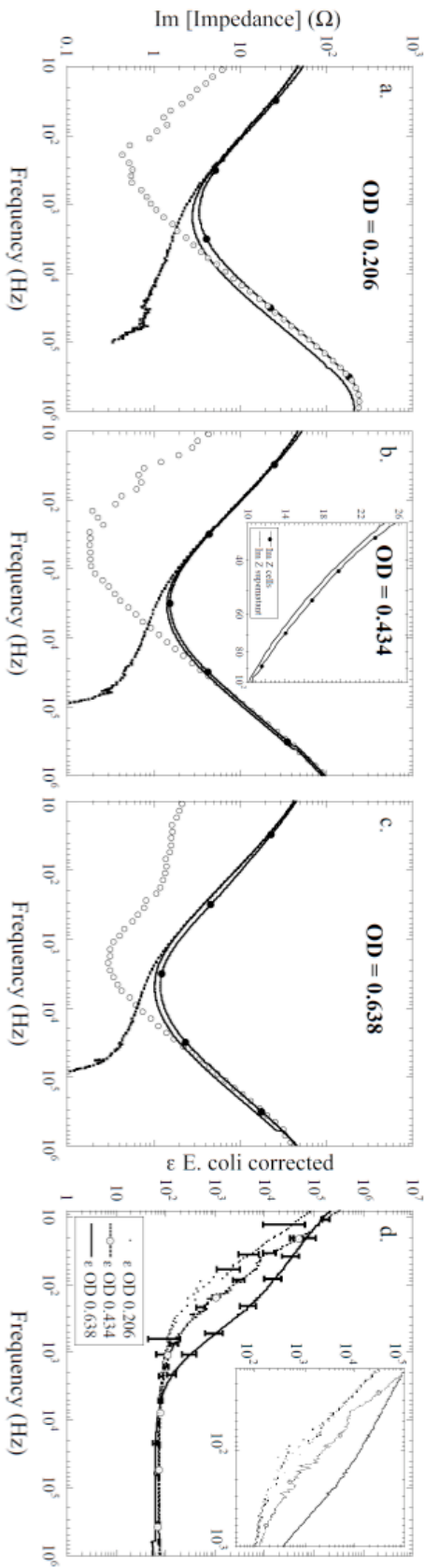

Fig 2 Bot et al 2008

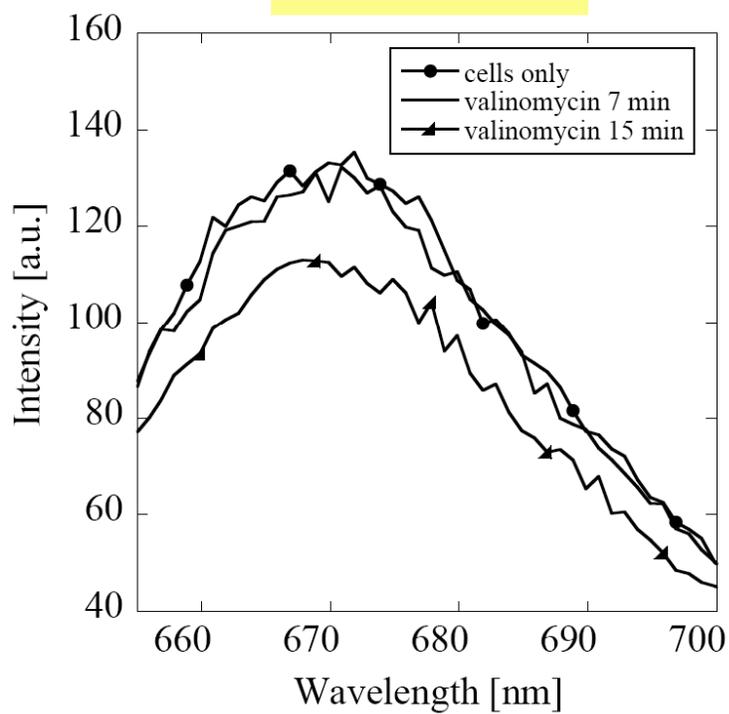

Fig. 3 Bot et al 2008

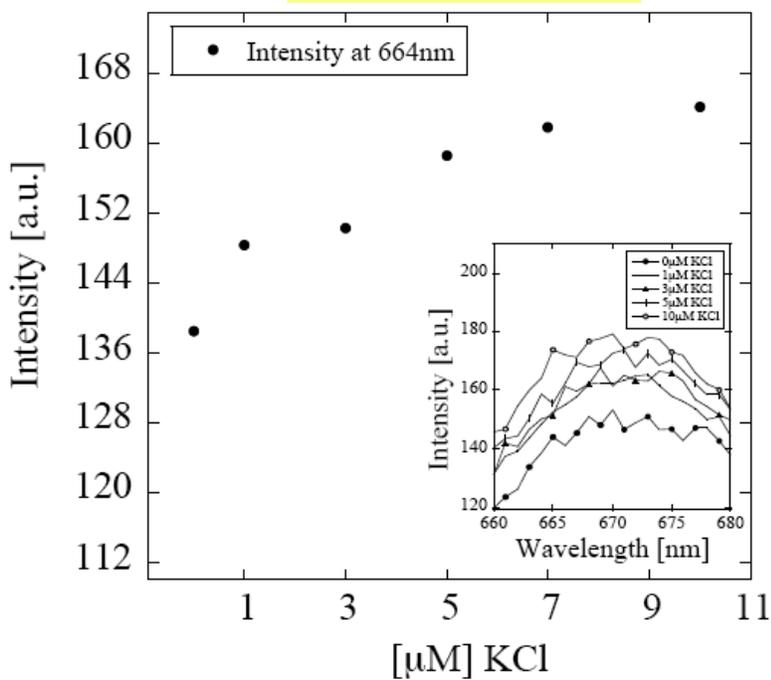

Fig. 4 Bot et al 2008

Fig. 5 Bot et al 2008

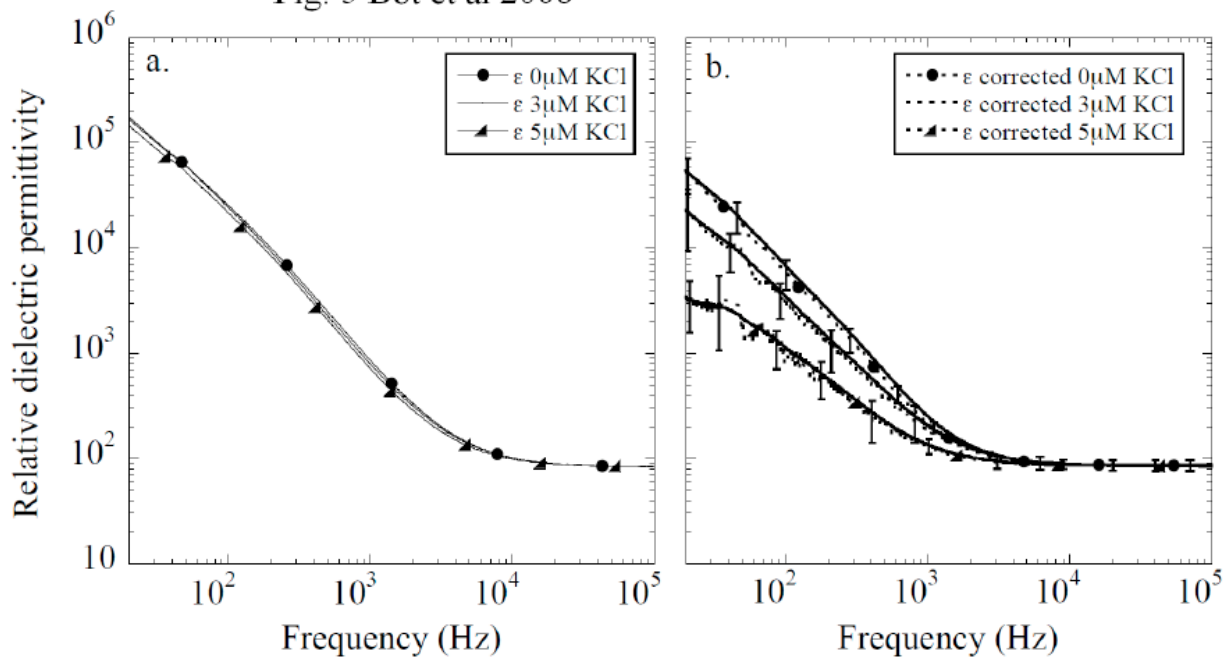

Fig. 6 Bot et al 2008

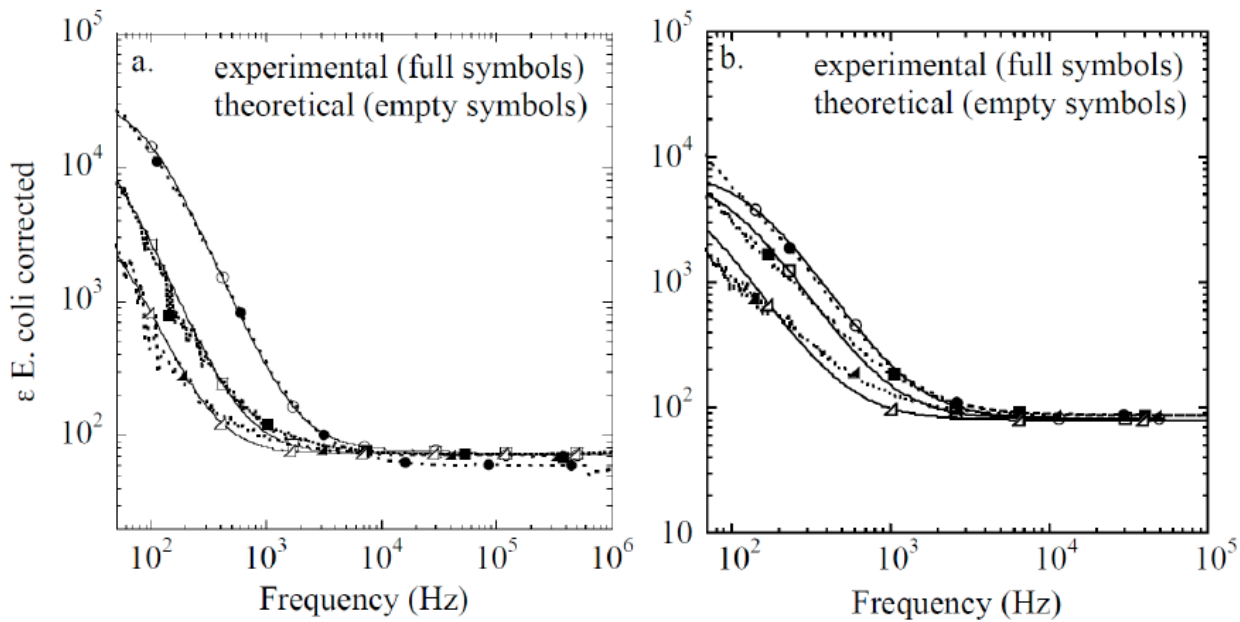

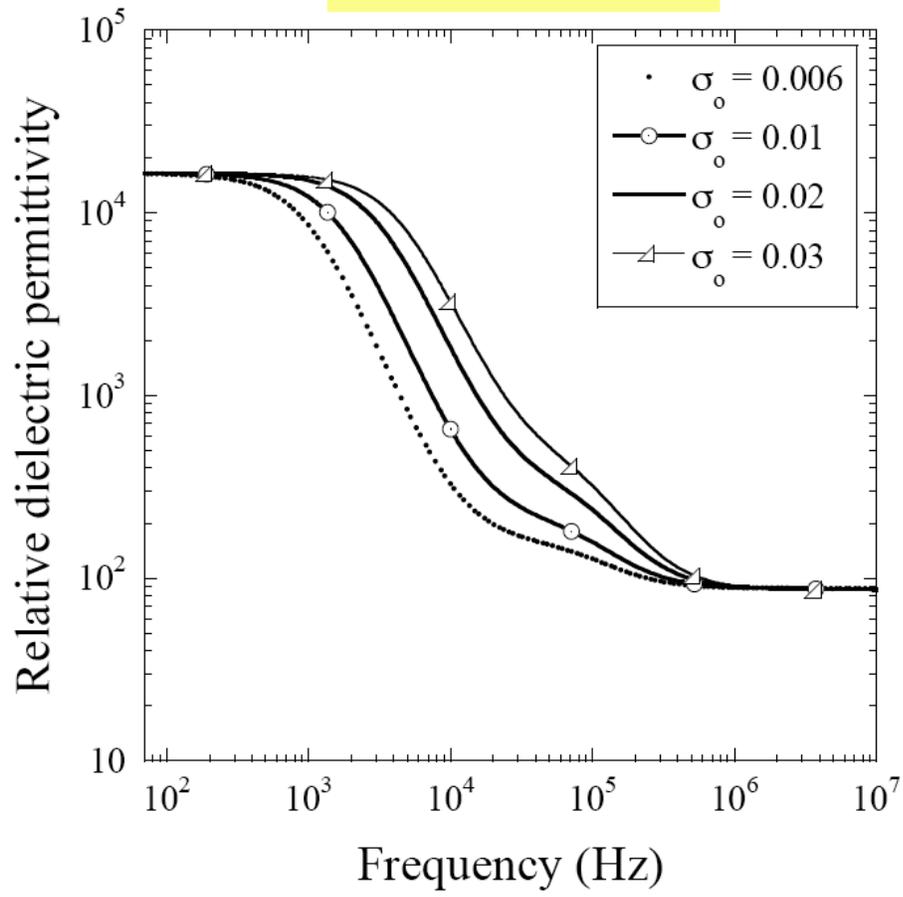
Fig. 7 Bot et al 2008